\documentclass[aps,prd,eqsecnum,nofootinbib,twocolumn,superscriptaddress,floatfix]{revtex4}
\usepackage{graphicx}
\usepackage{dcolumn}
\usepackage{times,mathptm}
\usepackage{subfigure}
\usepackage{psfrag}
\newcommand{\beq}{\begin{equation}}
\newcommand{\eeq}{\end{equation}}
\newcommand{\bea}{\begin{eqnarray}}
\newcommand{\eea}{\end{eqnarray}}

\renewcommand{\l}{\lambda}

\renewcommand{\b}{\beta}
\renewcommand{\a}{\alpha}

\newcommand{\g}{\gamma}
\newcommand{\n}{\nu}
\newcommand{\m}{\mu}
\renewcommand{\r}{\rho}
\newcommand{\bx}{{\mathbf{x}}}
\newcommand{\by}{{\mathbf{y}}}

\newcommand{\s}{\sigma}

\newcommand{\tS}{\widetilde{S}}

\newcommand{\tm}{\widetilde{\m}}

\newcommand{\oh}{{\textstyle{\frac{1}{2}}}}

\newcommand{\dg}{\dagger}
\newcommand{\non}{\nonumber}

\newcommand{\rf}[1]{(\ref{#1})}
\newcommand{\ra}{\rightarrow}
\newcommand{\pa}{\partial}

\begin{document}

\title{Is Confinement a Phase of Broken Dual Gauge Symmetry?}

\author{J. Greensite}
\affiliation{Physics and Astronomy Department., San Francisco State
University, San Francisco, CA~94132, USA}

\author{B. Lucini}
\affiliation{Department of Physics, Swansea University, Singleton Park, Swansea,
SA2 8PP, UK}

\date{\today}
\begin{abstract}
We study whether broken dual gauge symmetry, as detected by a monopole
order parameter introduced by the Pisa group, is necessarily 
associated with the confinement phase of a lattice gauge theory.  
We find a number of examples, including SU(2) gauge-Higgs theory, mixed 
fundamental-adjoint SU(2) gauge theory, and pure SU(5) gauge theory, which appear
to indicate a dual gauge symmetry transition in the absence of a transition to or from
a confined phase.  While these results are not necessarily fatal
to the dual superconductor hypothesis, they may pose some 
problems of interpretation for the present formulation of the Pisa monopole
criterion. 

\end{abstract}

\maketitle
%
%
\section{Introduction}\label{Intro}
 
     It is an old idea that the vacuum state of a non-abelian gauge theory in the confined phase can be thought of as a 
dual superconductor, with magnetic monopoles playing the role of Cooper pairs. Magnetic 
monopoles, in this well-known scenario, condense in the confined
phase, thereby spontaneously breaking a dual gauge symmetry, and electric flux tubes form as a consequence of the dual 
Meissner effect.  In testing this idea numerically, it is important to carefully specify what one means by monopole condensation, 
particularly in the absence of elementary Higgs fields, and to identify precisely which dual gauge symmetry is under consideration.

    It is worth recalling that the whole notion of spontaneously broken gauge symmetries, dual or otherwise, can be a little 
ambiguous.  According to the well-known Elitzur theorem \cite{Elitzur} a continuous gauge symmetry cannot break 
spontaneously, and the VEV of a Higgs field must vanish, regardless of the form of the Higgs potential, in the absence
of gauge fixing. For this reason, spontaneous gauge symmetry breaking must necessarily refer to the breaking of a global 
subgroup of the local symmetry, such as, e.g., a global symmetry which remains after fixing to Coulomb or Landau gauge. 
In principle there are an infinite number of such subgroups, and they do not necessarily break at the same place in the phase
diagram. Nor do such transitions \emph{necessarily} indicate a genuine change of phase.   In a recent article \cite{Billy} which considered
spontaneous gauge symmetry breaking in an SU(2) gauge-Higgs theory, it was shown that remnant gauge symmetries in the
Coulomb and Landau gauges do, in fact, break along different lines in the gauge-Higgs coupling space; moreover, these transitions
occur in regions where the Fradkin-Shenker-Osterweiler-Seiler theorem \cite{FS,OS} assures us that there is no transition whatever in the
physical state.  In view of this result, we think it may be of interest to revisit the issue of dual gauge symmetry breaking in pure gauge
theories, as formulated concretely by the Pisa group in refs.\ \cite{Adriano1,Adriano1a,ds1-2,Adriano2}. 

    The Pisa proposal is based on the fact that in a compact U(1) gauge theory there are stable magnetic monopole configurations,
and a corresponding conserved magnetic current
\beq
             j^M_\m = \pa^\n \widetilde{F}_{\m\n}
\eeq   
where
\beq
            \widetilde{F}_{\m\n} = \oh \epsilon_{\m\n\a\b}F^{\a\b}
\eeq
is the dual field strength tensor.  This conserved current is associated with a dual U(1) gauge symmetry, and a global subgroup of this
local symmetry is generated by the total magnetic charge operator.  An order parameter $\mu$ for the breaking of this global dual gauge symmetry
was proposed in refs.\ \cite{Adriano1,Adriano1a}.    The $\mu(\bx)$ operator is a monopole creation operator, which acts on states in the Schrodinger
representation by inserting a monopole field configuration $A_i^M(\by)$ centered at the point $\by=\bx$, i.e.
\beq
              \m(\bx) |A_i \rangle = | A_i+ A_i^M \rangle 
\label{pisa-op}Ä
\eeq
Explicitly, the operator
\beq
               \m(\bx) = \exp\Bigl[i \int d^3 y ~ A_i^M(y) E_i(y) \Bigr]
\label{m-charge}
\eeq
performs the required  insertion.   It was shown in ref.\ \cite{Adriano1} that the $\m$ operator is non-invariant under the global subgroup of
the dual gauge symmetry generated by the magnetic charge operator, and $\langle \mu \rangle \ne 0$ is the sign that this global symmetry
is spontaneously broken, i.e.\ that the system is in a phase of ``dual superconductivity".

    In a non-abelian SU(N) gauge theory, an abelian projection gauge  
must be introduced to single out an abelian $U(1)^{N-1}$ subgroup, and $\m$ is defined in terms of the
gauge fields associated with that subgroup.  In practice the choice of abelian projection gauge does not seem to make much difference; maximal abelian gauge, a Polyakov line gauge, and even a ``random" abelian gauge have all been tested, with very similar results.  Details concerning this construction on the lattice can be found in refs.\ \cite{ds1-2,Adriano2}.  

    For a lattice pure gauge theory, it can be shown analytically that $\langle \mu \rangle =1$ in the lattice coupling $\b \ra 0$ limit, and hence
the global dual gauge symmetry is spontaneously broken in that limit.  However, $\langle \m \rangle$ cannot be computed analytically at weak couplings, and must instead be determined numerically, via lattice Monte Carlo simulations.  Since it is impractical, numerically, to compute 
$\langle \m \rangle$ directly, one instead computes the logarithmic derivative
\beq
\r = {\pa \over \pa \b} \log\langle \mu \rangle  = - \b^{-1} \left[ \langle S \rangle_S 
    - \langle S_M \rangle_{S_M} \right]
\eeq
In this expression $S$ is the usual Wilson action,  and $\langle \cdots \rangle_S$ indicates an evaluation with the usual
probability measure $\propto \exp[S]$.  $\langle S_M \rangle$ is a monopole-modified action, in which the timelike plaquettes of the Wilson action on a time-slice are modified by the monopole field (we refer to refs.\ \cite{ds1-2} for details of this construction), and  
$\langle \cdots \rangle_{S_M}$ denotes a VEV with probability measure proportional to $\exp[S_M]$.  Then
\beq
            \langle \m(\b) \rangle = \exp\left[ \int_0^\b d\b' ~ \r(\b') \right]
\label{mrho}
\eeq
If there is some $\b_{cr}$ where $\r(\b_{cr}) \ra -\infty$ in the infinite volume $V\ra \infty$ limit, such that the integral in the exponent diverges for $\b>\b_{cr}$, it means that $\langle \m(\b) \rangle =0$ at $\b>\b_{cr}$, i.e.\ that there is a transition from the broken to the unbroken phase of global dual gauge symmetry.     

    Previous investigations of the Pisa operator have focused mainly on the deconfinement transition, for gauge theories with various gauge groups, with and without dynamical fermions.  In every case the deconfinement transition coincides with a sharp negative peak in $\r(\b)$, which grows deeper with lattice volume, implying that $\langle \m \rangle > 0$ in the confined phase, and $\langle \m \rangle = 0$ in the deconfined phase.  The same sort of negative peak in $\r$ was found in U(1) gauge theory, in D=4 dimensions and zero temperature, at the transition from the confining to the massless Coulomb phase \cite{Adriano1}.  All of these results are consistent with the view that the confining phase is a phase of broken dual gauge symmetry.  
    
    We will now present several possible counter-examples to that view, and show that at zero temperature, in
\begin{enumerate}
\item SU(5) lattice gauge theory
\item SU(2) lattice gauge theory with a mixed fundamental + adjoint action
\item SU(2) lattice gauge theory with the standard Wilson action
\item SU(2) gauge-Higgs theory
\end{enumerate}
there is a large negative peak in $\r$, growing deeper with lattice volume, suggesting that there can be restoration of the dual gauge symmetry without a corresponding transition away from a confined phase.  In these computations we follow the Pisa construction of $\r$ described in refs.\ \cite{Adriano1,Adriano1a,ds1-2,Adriano2}, with the random abelian projection advocated in \cite{Adriano2}.
 
\section{Results}
   
\begin{figure}[t!]
{\includegraphics[width=9truecm]{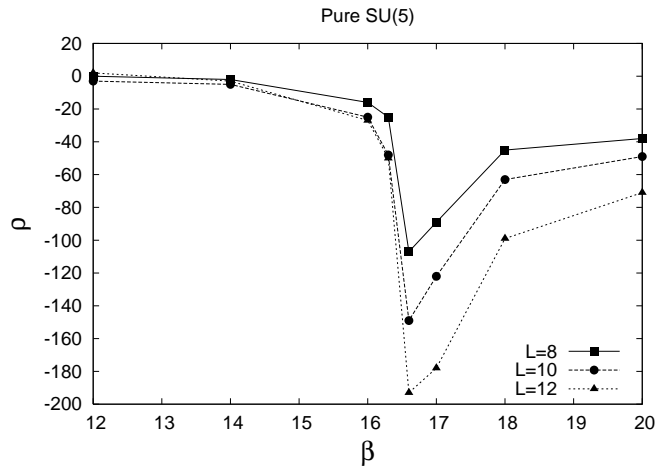}}
\caption{$\r$ vs.\ $\b$ in SU(5) lattice gauge theory with the standard Wilson action.  The negative peak occurs at the
bulk transition point.} 
\label{su5}
\end{figure}

    Our first example is SU(5) pure gauge theory, with the standard Wilson action, at zero temperature.   Many years ago it was found that there is a bulk first-order transition in this theory \cite{creutz}; the most recent determination of the transition point is at $\b=16.65$ \cite{biagio}.   In Fig.\ \ref{su5} we display our results for $\r$, which indicates a negative peak, growing with volume, at the bulk transition point.   The lattices are a bit too small to draw strong conclusions, but if the trend continues at larger volumes, then it would appear that $\langle \m \rangle = 0$ for $\b$ values beyond the bulk transition point, despite the fact that the bulk transition is not associated with a loss of confinement.

\begin{figure*}[t!]
\begin{center}
\subfigure[] 
{
    \label{mixact}
    \psfrag{b1}{$\b$}
    \psfrag{b2}{$\b_A$}
    \includegraphics[width=8truecm]{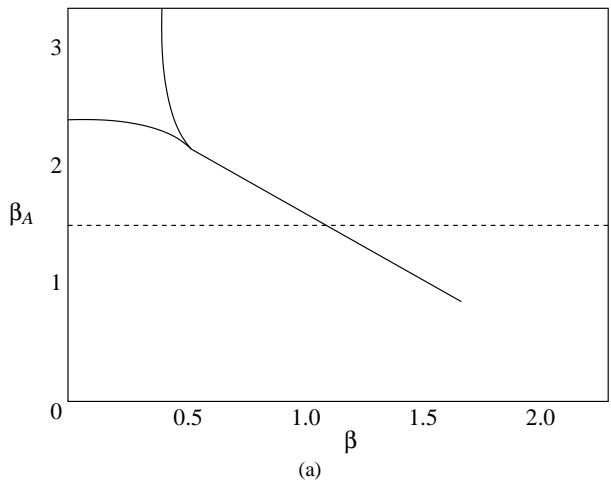}
}
\hspace{0.25cm}
\subfigure[] 
{
    \label{mix-rho}
    \includegraphics[width=8truecm]{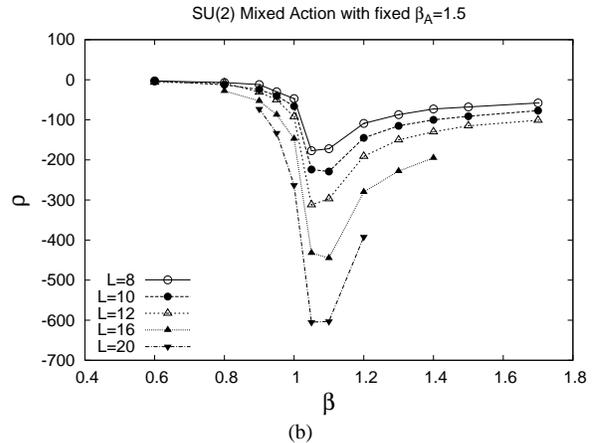}
}
\end{center}
\caption{(a) Schematic phase diagram of the SU(2) fundamental-adjoint mixed action model.  Our data points are taken along the
dashed line.  (b) $\r$ vs.\ $\b$ in the SU(2) fundamental-adjoint mixed action of eq.\ \rf{mix}, at $\b_A=1.5$.}
\label{ghiggs} 
\end{figure*}

\begin{figure}[t!]
{\includegraphics[width=9truecm]{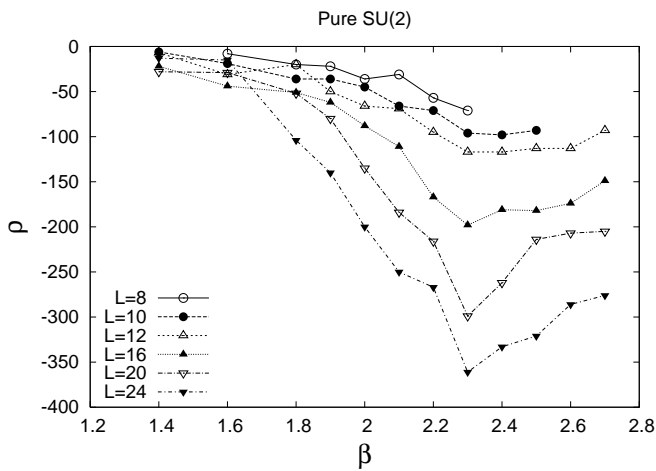}}
\caption{$\r$ vs.\ $\b$ in SU(2) lattice gauge theory with the standard Wilson action.} 
\label{su2}
\end{figure}

    The second example is the SU(2) fundamental-adjoint mixed action
\bea
           S &=& \b \sum_{plaq} \oh \mbox{Tr}_F[U(P)] + \b_A \sum_{plaq} {1\over 3} \mbox{Tr}_A[U(P)] 
\non \\
               &=&  S^{Wilson} + S^{Adjoint}           
\label{mix}
\eea
where Tr${}_{F,A}$ refer to traces in the fundamental and adjoint representations, respectively, of the SU(2) group.   The phase diagram of this theory was determined numerically, many years ago, by Bhanot and Creutz \cite{bhanot}; the diagram is drawn schematically in Fig.\ \ref{mixact}, and the solid lines indicate lines of first-order transition.    Keeping $\b_A$ fixed at $\b_A=1.5$, and varying only $\b$, we discover a negative peak in $\rho$, shown in Fig.\ \ref{mix-rho}.  In this case $\rho$ is given by
\beq
\r = {\pa \over \pa \b} \log\langle \mu \rangle  = - \b^{-1} \left[ \langle S^{Wilson} \rangle_S 
    - \langle S^{Wilson}_M \rangle_{S_M} \right]
\eeq
The location of this peak ($\b \approx 1.1$ at $\b_A=1.5$) lies on a line of first-order phase phase transitions.  Of course, the lines of first-order transitions in the $\b-\b_A$ coupling plane do not imply a deconfinement transition at zero temperature, since the mixed-action model is confining at all values of $\b,\b_A$.  However, the existence of a transition in 
$\langle \m \rangle$ does suggest that the dual gauge symmetry is restored in an entire region of the coupling plane, and it is of interest whether that region includes large $\b$ at $\b_A=0$; i.e.\ pure SU(2) lattice gauge theory with the usual Wilson action.  Fig.\ \ref{su2} shows 
our results for $\rho$ computed in ordinary SU(2) lattice gauge theory with the Wilson action.   In this data there does appear to be a very broad peak, growing with lattice volume, centered at about $\b=2.3$.  It is clear, in this case, that one really does need to go out to fairly large volumes, greater than $16^4$, to actually see the peak.  If we take this result at face value, then it means that the dual gauge symmetry is unbroken in the large $\b$ regime,
despite the fact that the theory is believed to confine at all values of $\b$.

\begin{figure}[b!]
\centerline{\rotatebox{270}{\includegraphics[width=6truecm]{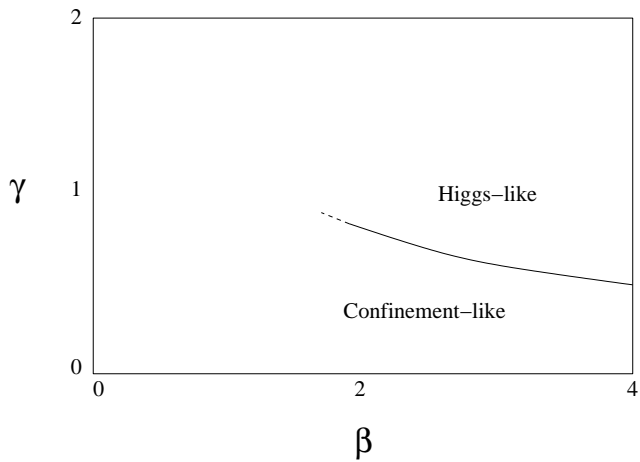}}}
\caption{Schematic phase diagram of the SU(2) gauge-Higgs system.  
The solid line is a line of either sharp crossover or weak first-order  
transitions.} 
\label{fsphase}
\end{figure}

    The next example is taken from the theory which motivated a previous investigation \cite{Billy} into the ambiguities associated with spontaneously broken gauge symmetries.  This is the ``Fradkin-Shenker" model, i.e.\ an SU(2) gauge-Higgs theory with a fixed modulus Higgs field in the fundamental representation of the gauge group, whose action can be written in the form
\bea
     S &=& \b \sum_{plaq} \oh \mbox{Tr}[U(P)] 
       + \gamma \sum_{x,\m} \oh \mbox{Tr}[\phi^\dg(x) U_\m(x) \phi(x+\widehat{\m})]  
\non \\
        &=& S^{Wilson} + S^{Higgs}  
\label{fradshenk}
\eea
where $\phi(x)$ is SU(2) group-valued.   What is important about this theory, for our purposes, is that it has been proven rigorously \cite{FS,OS} that there is no complete separation between the Higgs-like and confinement-like regions of the theory.  More precisely, consider a point $a$ in the $\b-\g$ phase diagram at $\b,\g \ll 1$ (confinement-like region), and another point $b$ at $\b,\g \gg 1$ (Higgs-like region).  Then the theorem states that there is a path between points $a$ and $b$ such that all correlators of all local gauge-invariant operators
\beq
             \langle A(x_1) B(x_2) C(x_3) \ldots \rangle
\eeq
vary analytically along the path.  This rules out any abrupt transition from a colorless to a color-charged spectrum.   So if by ``confinement" one simply means the absence of color-charged particles in the asymptotic spectrum, then the Higg's phase is ``confined" in this sense 
(c.f.\ the discussion in ref.\ \cite{Billy}).   Computer simulations \cite{campos} have found a phase structure indicated schematically in Fig.\ \ref{fsphase}.   The solid line is either a very weak first order transition, or a sharp crossover.  Campos, in ref.\ \cite{campos}, argues for a first-order transition, but the argument requires taking the limit of theories with varying modulus Higgs fields.  It is unimportant, for our purposes, whether the line represents genuine first-order transitions, or only crossover behavior. What \emph{is} important is that the Higgs-like and confinement-like regions are connected, and it is possible to go from one region to the other without encountering a physical transition of any sort.  The question we raise is whether this connectedness also holds true for the dual gauge symmetry, as detected by the $\m$ operator.  In this case $\m$ is a function of $\g,\b$, and it is convenient to define $\r$ in this model as the logarithmic derivative with respect to $\g$, i.e.\
\beq
\r = {\pa \over \pa \g} \log\langle \mu \rangle  = - \g^{-1} \left[ \langle S^{Higgs} \rangle_S 
    - \langle S^{Higgs} \rangle_{S_M} \right]
\eeq
so that
\beq
            \langle \m(\b,\g) \rangle = \langle \m(\b,0) \rangle \exp\left[ \int_0^\g d\g' ~ \r(\b,\g') \right]
\eeq
If there is only one phase, then one expects $\langle \m \rangle > 0$ everywhere in the phase diagram,
and there should not be any negative peaks, growing deeper with volume, in the $\rho$ observable.

\begin{figure*}[t!]
\begin{center}
\subfigure[] 
{
    \label{fs22}
    \includegraphics[width=8truecm]{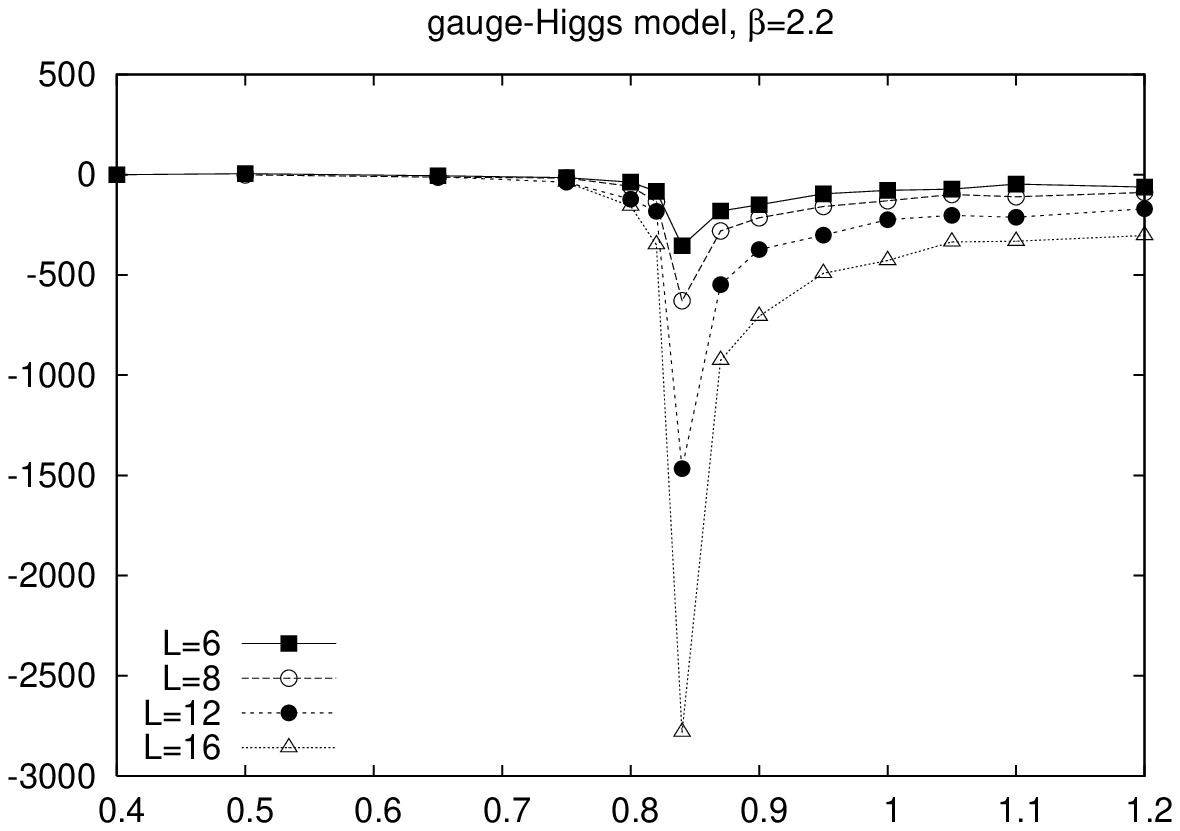}
}
\hspace{0.25cm}
\subfigure[] 
{
    \label{fs16}
    \includegraphics[width=8truecm]{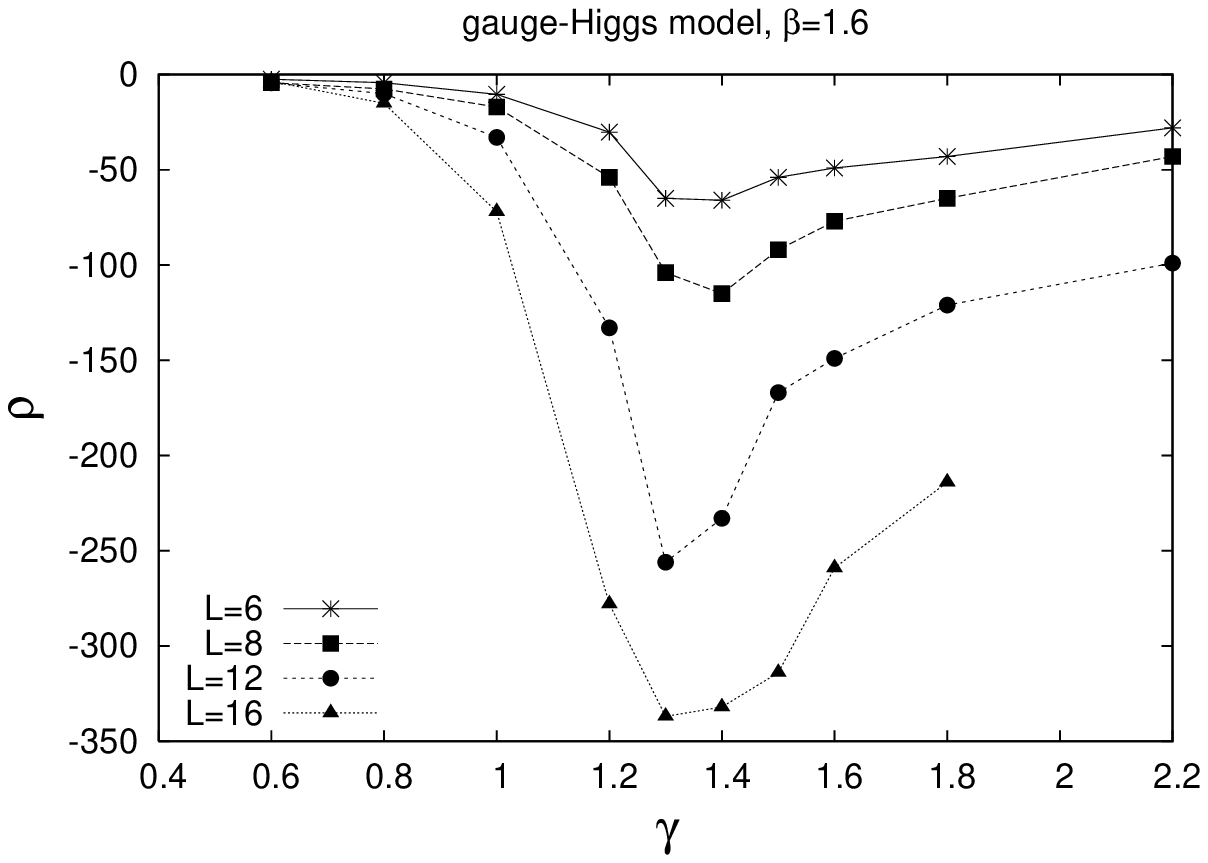}
}
\end{center}
\caption{$\r$ vs.\ $\b$ in the SU(2) gauge-Higgs (Fradkin-Shenker) model of eq.\ \rf{fradshenk}.  (a) At $\b=2.2$ a very
sharp negative peak in $\r$ is found at the thermodynamic transition/crossover point at $\g=0.84$.  (b) At $\b=1.6$, there is no thermodynamic transition, but nevertheless there is a broad negative peak in $\r$ centered at $\g \approx1.3$.}
\label{fs} 
\end{figure*}

    However, this expectation does not seem to hold.  In Fig.\ \ref{fs22} we show our data for $\r$ vs.\ $\g$ at $\b=2.2$, and we find a negative peak in $\r$ at $\g=0.84$, which coincides with a sharp crossover (or weak first-order transition) at this point.   Despite the fact that there is only one color neutral phase, the $\mu$ observable finds two phases.  But if there are two phases, corresponding to broken and unbroken symmetry, then the boundary between them must continue past the line of cross-over/weak first-order transitions.  In Fig.\ \ref{fs16}  we show our data for $\r$ vs.\ $\g$ at $\b=1.6$, where there is no physical transition of any kind.  However, there again appears to be a broad peak in $\r$, centered at $\g=1.3$, which is growing deeper with lattice volume.

     The final example is not due to us, but rather to Cossu et al.\ in ref.\ \cite{Cossu}.   These authors have computed $\r$ for the G(2) group, and found a sharp negative peak, growing deeper with volume, at a point of sharp crossover behavior found at $\b\simeq 1.35$.  A negative peak at the high-temperature deconfinement transition is also visible, but is actually very small in comparision to the negative peak at the crossover point.   
The existence of a transition to $\langle \m \rangle = 0$ at $\b\simeq 1.35$ in G(2) lattice gauge theory, where there is no actual transition to or from a color-neutral spectrum, fits neatly together with the other examples we have shown above.
 
\section{Discussion}
    
     All of the data shown above suggests that the $\m$ operator, as currently defined, has zero-temperature transitions at couplings where there are, in fact, no \emph{physical} transitions to or from a confined phase.  We have found sharp negative peaks in $\r$ where there are bulk transitions or crossover points in the phase diagram, and also much broader peaks, which show up even in the absence of any activity in the thermodynamics. Although a finite size scaling analysis of $\rho$ goes beyond the scope of this paper, our numerical data are clearly compatible with a divergence of the peaks in the infinite volume limit and seem to exclude sharp crossovers in all the cases we have investigated.  It is then reasonable to infer from this behavior that $\langle \m \rangle = 0$ past the peak in the infinite volume limit.  There are then two possibilities:  Either the $\m$ operator is an unreliable indicator for the presence of broken or unbroken dual gauge symmetry, i.e.\ we can have $\langle \m \rangle = 0$ even in the broken phase, or else confinement is not necessarily a phase of spontaneously broken dual gauge symmetry.  A third, less likely alternative is that the negative peaks in $\r$ cease to grow deeper beyond some
finite volume.  Such a possibility can never be ruled out just from numerical simulations.  However,  a very rough estimate from our data is that this convergence in $\r$, if it were to occur, would have to happen at distance scales beyond 4-5 fm, corresponding to 40-50 Mev.  The origin of such a low energy scale, in the theories we have considered, is rather unclear.
     
     If $\m$ is the wrong operator to use to test for spontaneous breaking of dual gauge symmetry, then one would like to understand exactly 
\emph{why} it is the wrong operator.   In particular it is important to explain why $\langle \m \rangle = 0$ when the dual gauge symmetry is broken,
and to construct a better operator $\tm$ which shows a transition if and only if there is a transition to or from the broken symmetry phase.   It may well be possible to construct such an operator.  However, the following remarks may be relevant to certain approaches: \\

\subsection{Rescale $\m$}

  One could easily construct a variant of $\m$ which avoids bulk transitions in SU(5) and G(2) (along the lines of a suggestion in ref.\ \cite{Cossu}) by normalizing the $\m$ operator at finite temperature $T$  by its value at zero temperature.   Since $\langle \m \rangle$ at any finite volume depends on the temperature $T$, the  volume of a time-slice $V_{D-1}$, and lattice coupling $\b$, we might, e.g., define a renormalized operator
\beq
           \tm(T,\b) \equiv \lim_{T_0 \ra 0} \lim_{V_{D-1}\ra \infty} {\m(T,\b,V_{D-1}) \over \langle \m(T_0,\b,V_{D-1}) \rangle}
\label{renorm}
\eeq
In practice, one might redefine $\r$ at at any temperature by subtracting its value at zero temperature, and computing $\langle \tm \rangle$ from eq.\ \rf{mrho} with the subtracted $\r$.  
An operator defined in this way is obviously insensitive to the bulk transition, but it may be sensitive to transitions at the deconfinement temperature.   However, this redefinition begs the question of whether the dual gauge symmetry is realized in the broken or unbroken phase at high $\b$ and zero temperature.  An operator which is, by definition, unity at zero temperature for all $\b$ and all gauge groups, in all dimensions $D$, clearly cannot address this crucial issue.

\subsection{Modify $S$}

    Bulk transitions can also be avoided, in certain cases, by modifying the lattice action.   For example, SO(3) lattice gauge theory (set $\b=0$ in the mixed action of eq.\ \rf{mix}) is known to have a bulk transition at $\b_A \approx 2.5$, and the transition is associated with certain lattice artifacts known as $Z_2$ monopoles.  The bulk transition can be avoided by introducing a term in the action which suppresses these monopoles
\beq
         \tS =  \b_A \sum_{plaq} {1\over 3} \mbox{Tr}_A[U(P)] + \lambda \sum_{cubes} \s_c
\label{tS}
\eeq
where 
\beq
     \s_c = \prod_{P \in \pa c} \mbox{sign}(\mbox{Tr}[U(P)]
\eeq
and the product runs over all plaquettes $P$ on the boundary of cube $c$.\footnote{Data on the Pisa operator at finite temperature in this model can be found in ref.\ \cite{MMP}.}  At sufficiently large $\l$ the bulk transition disappears.   Perhaps some modification of the lattice action would also avoid bulk transitions in other theories.  

    However, it does \emph{not} follow that eliminating a bulk transition via a modified action will also eliminate the transition to $\langle \m \rangle \ra 0$ at large $\b$ and zero temperature.  We assume that the modified and unmodified actions have the same continuum limit, in the sense that expectation values
\beq
            \langle Q \rangle_{\tS} \ra \langle Q \rangle_S  ~~~ \mbox{at large~} \b
\label{limit}
\eeq
But if that is the case, and $\langle \m \rangle_S = 0$ for the unmodified action at large $\b$, then if \rf{limit} holds it must also be true that 
$\langle \m \rangle_{\tS} = 0$ at large $\b$ in the modified action.  Of course, the logarithmic derivative $\r$ is different for $S$ and $\tS$, and it may be that the transition from $\langle \m \rangle_{\tS} \ne 0$ to $\langle \m \rangle_{\tS} = 0$ is harder to detect on finite lattice volumes with the modified action.  In fact, we have seen in two examples
\begin{enumerate}
\item pure SU(2) gauge theory with the Wilson lattice action at $\b=2.3$;
\item the Fradkin-Shenker model at $\b=1.6$ and $\g=1.3$
\end{enumerate}
that there are transitions in $\m$ in the absence of any thermodynamic transition or crossover behavior.   However, in these cases the peaks are much broader than in our other examples where the transitions coincide with thermodynamic activity.  Especially in the case of pure SU(2) lattice gauge theory, the transition in  $\m$ is only convincingly seen at larger lattice volumes.  This may also be the case when using a modified action $\tS$ which eliminates the bulk transition, but still has the property \rf{limit}.  It is certainly possible that the transition in $\langle \m \rangle_{\tS}$ happens at a different $\b$ than
the transition in $\langle \m \rangle_{S}$, and is only seen clearly at very large lattice volumes.  But whether or not the peak in $\r$ is readily seen at small volumes with the modified action, a transition in  $\langle \m \rangle_{S}$ \emph{implies}
a transition in $\langle \m \rangle_{\tS}$.    In a sense, the fact that $\r$
has a prominent negative peak at the bulk transition point of the unmodified action is fortunate, because it informs us of the existence of a dual gauge symmetry restoration at weak couplings, that might otherwise have been missed.

\subsection{Modify $\m$}

   Perhaps there is some other way to modify $\m$, apart from the rescaling \rf{renorm}, such that the modified observable $\tm$ would remain non-zero across bulk and other unphysical transition points.  We cannot say whether this is possible or impossible, but only remark that if $\tm$ approaches $\m$ in the continuum limit, i.e.\ if
\beq
             \langle \tm \rangle_S \ra \langle \m \rangle_S   ~~~ \mbox{at large~} \b
\label{mlimit}
\eeq
then, for same reasons as before, $\langle \m \rangle = 0$ at large $\b$ implies $\langle \tm \rangle = 0$ at large $\b$.  Once again, the transition from $\langle \tm \rangle \ne 0$ to $\langle \tm \rangle = 0$ may occur at a different value of $\b$ than the transition in $\langle \m \rangle$, and the peak may be much broader and harder to detect at small volumes, but the limiting behavior \rf{mlimit} is essentially a guarantee that the transition 
must exist at some coupling. \\\

\section{Conclusions}

    The trend of our data suggests that either the Pisa $\m$ operator, as currently defined, is not a reliable order parameter for the dual gauge symmetry breaking, or else that the confinement phase of a gauge theory is not necessarily a phase in which that symmetry is broken.   If $\m$ is not a reliable order parameter, then it is necessary to understand why this is so; i.e.\ how it can happen, in a phase of broken symmetry, that nevertheless one may find $\langle \m \rangle = 0$, even at weak lattice couplings where lattice artifacts (such as $Z_2$ monopoles) are presumably irrelevant.  If this point can be understood, then perhaps a more viable order parameter can be constructed.

    The other possibility is that confinement is not \emph{necessarily} tied to the breaking of a dual gauge symmetry, and it is possible for confinement to exist in the unbroken phase. In fact this would fit in rather well with previous results reported in ref.\ \cite{Billy}.   There it was pointed out that different global subgroups of a gauge symmetry, associated with the Kugo-Ojima confinement criterion \cite{Kugo} and the Coulomb confinement criterion \cite{coulomb} respectively, break along different lines in the Fradkin-Shenker phase diagram, in regions where there does not exist any physical change of phase.  Thus the very concept of a spontaneously broken gauge symmetry is a little ambiguous (since a global subgroup must be specified),  and the broken or unbroken realization of the symmetry is not necessarily tied to confinement.  The latter conclusion may also hold for global subgroups of the dual gauge symmetry.   
    
    At this stage, we think it is not yet clear whether confinement is really independent of dual gauge symmetry breaking, or whether, alternatively, $\m$ as currently formulated is simply an imperfect order parameter for that breaking.  The issue is not settled, and will require some further investigation.

\acknowledgments{We have benefited from discussions with Guido Cossu, Massimo D'Elia and Adriano di Giacomo.
B.L.\ is supported by a Royal Society University Research Fellowship; J.G.\ is supported in part by  
the U.S.\ Department of Energy under Grant No.\ DE-FG03-92ER40711.}

\end{document}